\documentclass{iopjournal}

\usepackage{graphicx}  
\graphicspath{{figures/}}
\usepackage{amsmath}
\usepackage{amssymb}
\usepackage{tikz}
\usetikzlibrary{shapes.geometric, arrows}

\tikzstyle{startstop} = [rectangle, rounded corners, minimum width=1cm,
  minimum height=0.5cm, text centered, draw=black, fill=red!30]
\tikzstyle{io} = [trapezium, trapezium left angle=70,
  trapezium right angle=110, minimum width=1cm, minimum height=0.5cm,
  text centered, draw=black, fill=blue!30]
\tikzstyle{process} = [rectangle, minimum width=1cm, minimum height=0.5cm,
  text centered, draw=black, fill=orange!30]
\tikzstyle{decision} = [diamond, minimum width=0.5cm,
  minimum height=0.5cm, text centered, draw=black, fill=gray!30]
\tikzstyle{arrow} = [thick,->,>=stealth]

\newcommand{\pone}{p_1} 
\newcommand{\azero}{a_0}  
\newcommand{\xof}{x_{\mathrm{off}}}  
\newcommand{\cNtwo}{c_{\mathrm{N_2}}} 
\newcommand{\Emed}{E_{\mathrm{med}}}  
\newcommand{\dEmad}{\delta E_{\mathrm{mad}}} 
\newcommand{\Q}{Q}  
\newcommand{\Rsq}{R^2}
\newcommand{\thetastar}{\theta^{\star}}       
\newcommand{\bstar}{b^{\star}}

\begin{document}

\articletype{PAPER}

\title{Inverse modelling of a laser-plasma accelerator}

\author{M. Lenivenko$^{1,*}$\orcid{0009-0009-8753-8711},
        A. Beck$^2$\orcid{0000-0002-8629-6812},
        H. G\"uler$^1$\orcid{0000-0002-9709-8715},
        G. Kane$^1$\orcid{0009-0004-9368-0734},
        S. Kazamias$^1$\orcid{0009-0008-8266-1692},
        V. Kubytskyi$^1$\orcid{0000-0000-0000-0000},
        B. Lucas$^1$\orcid{0009-0000-4070-828X},
        M. Pintonato$^1$,
        J. Serhal$^1$\orcid{0009-0002-5219-7653},
        S. Smartsev$^1$\orcid{0000-0002-6445-6184},
        K. Swain$^1$\orcid{0009-0005-9716-1234},
        and K. Cassou$^1$\orcid{0000-0002-6061-0447}}

\affil{$^1$Laboratoire de Physique des 2 Infinis Irene Joliot-Curie - IJCLab -
UMR9012 - Bat. 100 - 15 rue Georges Clemenceau 91405 Orsay cedex - France.}

\affil{$^2$Laboratoire Leprince-Ringuet - LLR - UMR 7638 CNRS Ecole
polytechnique, 91128 Palaiseau cedex - France.}

\affil{$^*$ Corresponding Author.}

\email{mykyta.lenivenko@ijclab.in2p3.fr}


\begin{abstract}
The electron beams of a laser-plasma accelerator depend non-linearly on the laser and gas settings. This dependence is explored using particle-in-cell (PIC) simulations, and machine-learning models trained on these simulations predict the beam from the settings. Here we address the reverse problem: given a measured beam, which settings produced it? We infer three of them: the gas pressure $\pone$, the normalised vector potential $\azero$ and the nitrogen concentration $\cNtwo$. The solution is not guaranteed to be unique. We train an ensemble of mixture density networks that, for each beam, output several possible sets of settings with their probabilities and uncertainties, and test them on simulated beams with realistic measurement noise. On average, the predictions are accurate ($\Rsq = 0.91$), yet for about two-thirds of the beams the inferred distribution has two competing solutions. These share nearly the same value of $\azero^2\pone$, which controls how strongly the laser focuses itself in the plasma and, in earlier PALLAS studies, when electrons start to be injected. Keeping this quantity fixed gives beams that are practically indistinguishable with realistic diagnostics. A forward model used as a check removes unphysical solutions but cannot choose between the two plausible ones. We then test two ways to resolve the ambiguity. Adding the full electron energy spectrum improves the mean prediction ($\Rsq = 0.94$), mostly through $\azero$. It changes the weights of the two solutions but does not move them. A second measurement at a deliberately different plasma density is more effective. In paired PIC simulations, it reduces the error on $\azero$ by a factor of $2.6$, whereas repeating the measurement at the same density brings almost no gain. 
\end{abstract}

\section{Introduction}\label{sec:intro}
In a laser-plasma injector (LPI), the measured electron bunch is the end point of a coupled laser-plasma interaction, and it carries information about the effective conditions in the plasma that produced it. These effective conditions can differ from the recorded set points due to jitter, drift, and calibration offsets. Inferring them from the beam itself provides an independent, shot-by-shot check of the machine state. We ask which laser and plasma inputs are compatible with an observed beam, and whether that answer is unique. An inverse model that answers this is useful for diagnosing drifts, checking shot records and translating simulated working points into experimental settings.

Laser wakefield acceleration (LWFA) provides compact electron sources with large accelerating fields and femtosecond bunch durations~\cite{tajima1979laser,esarey2009physics,faure2004laser}. Experiments have demonstrated high-energy beams in this regime~\cite{gonsalves2019petawatt}. The same non-linear dynamics that make LWFA compact also make the source difficult to tune. Several laser and plasma parameters simultaneously affect the final energy, charge, energy spread, emittance and duration. A small change in one input parameter can be compensated by other parameters, so the map from inputs to diagnostics is not one-to-one.

Particle-in-cell (PIC) simulations remain the primary tool for exploring these couplings, but they are too expensive to use directly in routine optimisation. Machine-learning surrogate models are now a common tool for accelerator design~\cite{dopp2023data}. They have been used for Bayesian optimisation and tuning of LWFA working points~\cite{jalas2021bayesian,jalas2023tuning,ferranpousa2023bayesian,irshad2024pareto} and for data-driven modelling of complete accelerator-based sources~\cite{brogren2025datadriven}. For the LPI considered here, the forward surrogate models trained on PIC simulations showed that beam characteristics can be predicted from input parameters with $\Rsq > 0.9$~\cite{kane_direct_model,drobniak2023random}.

In the present study, we work in the inverse direction. A forward surrogate provides a fast prediction of the beam for a proposed set of inputs. However, it does not indicate whether a measured beam corresponds to a single input setting or to several settings that would produce an indistinguishable electron beam. If several input configurations give the same beam within diagnostic accuracy, a single-point inverse model can return an average over valid settings. That average may not correspond to a physically useful working point. The inverse model should return all candidate solutions.

This posterior view of the laser-plasma inverse problem has been explored before. An invertible neural network trained on particle-in-cell simulations has been used to learn the forward map and to reconstruct simulation parameters from synthetic electron energy spectra~\cite{bethke2021invertible}, and follow-up work has benchmarked such networks on degenerate spectra from near-critical plasmas~\cite{miethlinger2023acceptance}. These studies invert idealised simulated spectra. The present work differs from them in both its inputs and its treatment of uncertainty. The inverse model uses scalar beam features and a per-channel measurement-noise model that reflects realistic diagnostic behaviour. We report how often the predictive intervals contain the true parameters, and rescale them so that this frequency matches the level they claim. The dominant ambiguity is traced to a combination of the input parameters, $\azero^2\pone$, that the forward map leaves almost unchanged: shots differing along this direction produce nearly indistinguishable beams. We then test which additional measurements break this near-invariance.

We construct such an inverse model using an ensemble of mixture density networks (MDN)~\cite{bishop1994mdn}. The model maps noisy beam characteristics and the assumed-known focal-position parameter to a posterior distribution over the remaining LPI inputs. We train and test it on the PALLAS PIC datasets~\cite{kane_direct_model,drobniak2023random}. We then use the predicted mixture components as candidate physical modes and pass them through a forward model to test which candidates reproduce the desired beam.

Sections~\ref{sec:setup} and~\ref{sec:methods} describe the datasets, the noise model and the MDN ensemble. Section~\ref{sec:accuracy} establishes the accuracy and calibration of the posterior, and section~\ref{sec:degeneracy} identifies the dominant degeneracy of the inverse map as the approximate invariant $\azero^2\pone$. Section~\ref{sec:reducing} tests the electron spectrum, a forward surrogate used as a referee, and a second measurement at a known plasma density as ways to reduce degeneracy. Section~\ref{sec:discussion} discusses the limitations and the measurements that would make the inverse solution identifiable.

\section{Setup: simulations and data}\label{sec:setup}

\subsection{Particle-in-cell simulation scans}\label{sec:setup:sims}

\begin{figure*}[h]
  \centering
  \includegraphics[width=\textwidth]{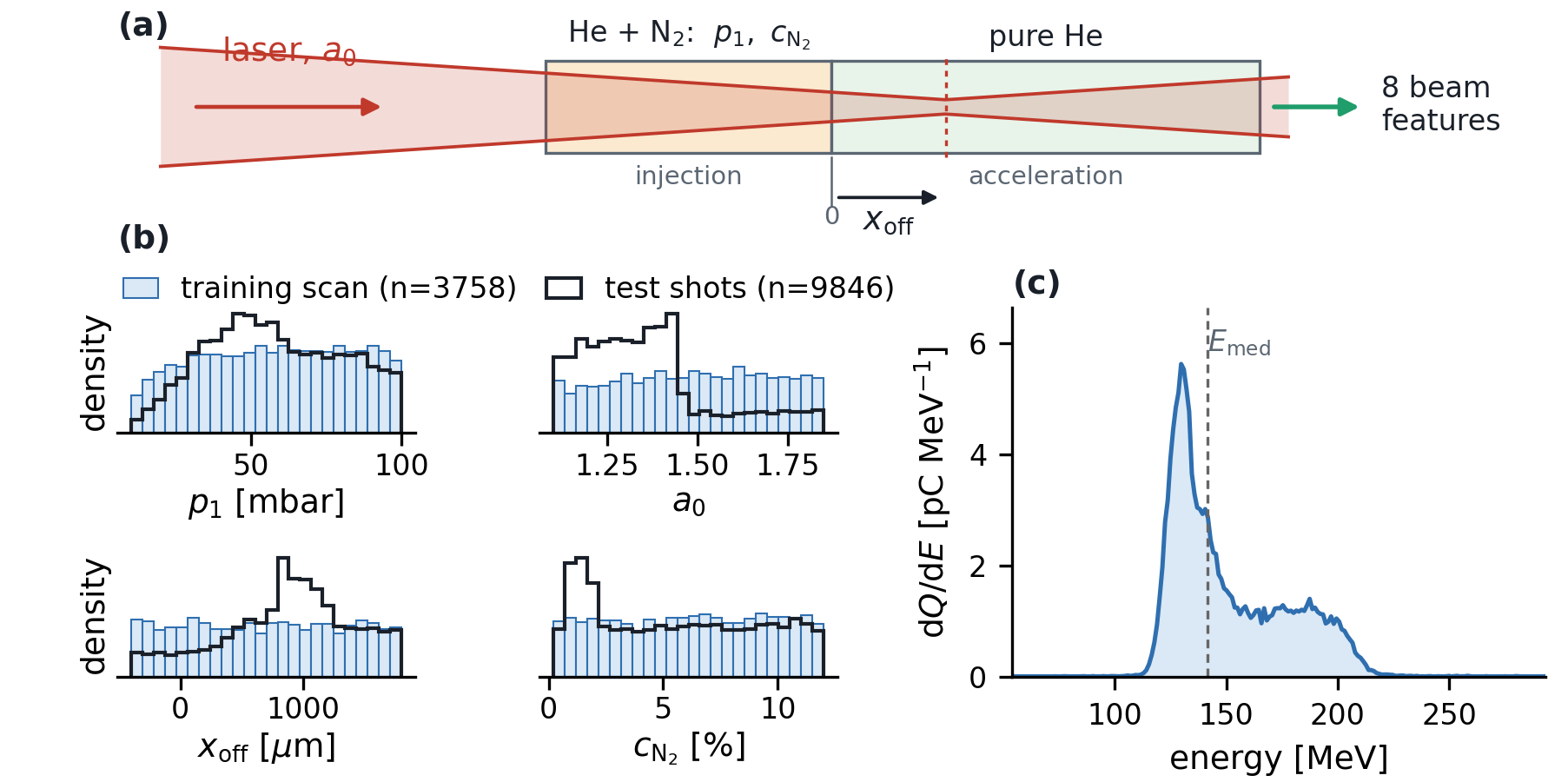}
  \caption{(a) The laser-plasma injector and its four input parameters. The laser with vacuum strength $\azero$ is focused at $\xof$ into a gas target with two regions, a first region filled with a He-N$_2$ mixture at pressure $\pone$ and nitrogen fraction $\cNtwo$ where electrons are injected, and a second region of pure helium where they are accelerated. The eight beam features of section~\ref{sec:setup:features} are measured at the exit. (b) Distribution of the four input parameters for the uniform training scan ($n_{\mathrm{train}} = 3758$, filled) and the injected random-scan test shots ($n_{\mathrm{test}} = 9846$, line). Each histogram is normalised to unit area. (c) One simulated electron spectrum from the test set, the shot closest to the test-set median energy, energy spread and charge ($\Emed = 141$~MeV, $\Q = 170$~pC). The dashed line marks $\Emed$.}
  \label{fig:setup}
\end{figure*}

The training and evaluation data are the two PIC simulation datasets of the forward model study~\cite{kane_direct_model} and the LPI design study at PALLAS~\cite{drobniak2023random}. Simulations were produced with the \textsc{Smilei} code~\cite{derouillat2018smilei} using azimuthal mode decomposition and the laser envelope approximation \cite{massimo2019envelope,massimo2020cylindrical}. The LPI relies on an ionisation injection scheme with a two-region gas target: a first region filled with a He-N$_2$ mixture for injection, followed by a pure-He region for acceleration~\cite{drobniak2023random}. Each simulation is defined by four input parameters: the laser normalised vector potential in vacuum $\azero$, the laser focal position in vacuum $\xof$, the gas pressure in the first region $\pone$, in mbar, and the nitrogen concentration in the first region $\cNtwo$. All other laser and gas parameters are the same in all simulations. Figure~\ref{fig:setup}(a) sketches the target and marks where each of the four inputs acts. Figure~\ref{fig:setup}(b) shows the range of each input and how the training and test configurations are distributed inside it.

The \emph{training set} is a uniform random scan dataset: input configurations drawn from a continuous uniform distribution over the input parameter ranges and prefiltered with the injection-prediction model of the forward model study~\cite{kane_direct_model}. It contains $n_{\mathrm{train}} = 3758$ simulations that produce an injected electron bunch.

The \emph{test set} is the random-scan dataset from~\cite{drobniak2023random}. We use the $n_{\mathrm{test}} = 9846$ shots marked as injected in the stored dataset. This test-set definition is used throughout the article, as without injection, there is no beam to measure, and the inverse task is not defined. The uniform distribution of the training inputs was shown in the forward-model study to be the decisive factor in surrogate accuracy across the full parameter space~\cite{kane_direct_model}. We follow the same choice. All models in this article are trained on the uniform scan and evaluated on the injected random-scan shots, unless otherwise stated. Both datasets are available in the public release of the forward model study~\cite{lpisurrogate_data}.

\subsection{Inverse task and beam features}\label{sec:setup:features}

The inverse task is to infer \((\pone, \azero, \cNtwo)\), with predictive uncertainties, from the measured characteristics of a single electron bunch, together with the machine setpoint \(\xof\). We take \(\xof\) as a known but imprecise condition. A preliminary study included \(\xof\) among the inference targets and recovered it only partially from the beam features. Supplying the known set point as an input is more informative than inferring it. This is consistent with the degeneracy analysis of section~\ref{sec:degeneracy}, where \(\xof\) dominates the local null direction of the forward map.

The bunch is summarised by eight scalar beam features: the median energy $\Emed$, the fractional energy spread $\dEmad$ (based on the median absolute deviation), the charge $\Q$, the peak current $i_{\mathrm{peak}}$, the normalised emittance $\epsilon_{n,x}$ along the axis orthogonal to the laser polarisation, the bunch length $\sigma_z$, the transverse size $\sigma_y$ and the RMS divergence. The beam features are evaluated at the last time step of each simulation, after the plasma outramp~\cite{kane_direct_model} with the \textsc{APtools} post-processing of the particle output~\cite{aptools}.

The first six features were selected in a preliminary feature selection study. Starting from an empty input list, we added, at each step, the diagnostic that provides the largest improvement in validation accuracy under a 30\% noise model. The selection metric was the reduction in the mean squared error on the validation set for the inferred target parameters. The transverse size $\sigma_y$ and the divergence were added in a later optimisation round under lower-noise conditions. Together with $\xof$, this gives nine model inputs. Features with a large dynamic range are $\log_{10}$-transformed, and all inputs are standardised. The feature set includes the bunch length $\sigma_z$, which is difficult to measure directly on the PALLAS electron beamline. This is a simulation study, and section~\ref{sec:discussion} quantifies the cost of removing $\sigma_z$.

\subsection{Measurement-noise model}\label{sec:setup:noise}
Experimental diagnostics deliver the beam features with finite accuracy. To account for this, a per-channel noise model is applied to the simulated features during training (resampled every epoch, serving as data augmentation) and once to the test set. Each channel $x$ receives a random error $u$. The error is drawn uniformly between $-\ell$ and $+\ell$, where $\ell$ is the half-width of the channel. The beam features have a relative error, $x' = x\,(1 + u)$. A half-width $\ell = 0.10$ means an error of up to $\pm 10\%$. The focal position $\xof$ has an absolute error, $x' = x + u$, with $u$ uniform in $[-a, +a]$. Each perturbed value is clipped to the physical range of its channel. Table~\ref{tab:noise} lists $\ell$ or $a$ for each channel. The $\pm10\%$ accuracy of the reconstructed quantities is an assumption. Appendix~\ref{app:noise} rescales each channel by 0 to 4 times this value.

\begin{table}[h]
  \caption{Per-channel measurement-noise model (uniform half-widths). Channels marked GPSR are assumed to be obtained from a generative 6-D phase-space reconstruction~\cite{roussel2024gpsr}.}
  \label{tab:noise}
  \centering
  \small
  \begin{tabular}{llll}
    \hline
    Channel & Noise & Type & Assumed source \\
    \hline
    $\xof$                    & $\pm 100~\mu$m & absolute & machine set point \\
    $\Emed$                   & $\pm 3\%$  & relative & spectrometer \\
    $\dEmad$                  & $\pm 10\%$ & relative & spectrometer \\
    $\Q$                      & $\pm 10\%$ & relative & charge diagnostic \\
    divergence (RMS)          & $\pm 10\%$ & relative & screen \\
    $i_{\mathrm{peak}}$       & $\pm 10\%$ & relative & GPSR \\
    $\epsilon_{n,x}$          & $\pm 10\%$ & relative & GPSR \\
    $\sigma_y$                & $\pm 10\%$ & relative & GPSR \\
    $\sigma_z$                & $\pm 10\%$ & relative & GPSR \\
    \hline
  \end{tabular}
\end{table}

The noise levels are defined according to the diagnostics foreseen for the experiment. The median energy is assigned a smaller uncertainty than the charge, energy spread and divergence. The phase-space quantities are assumed to be obtained from a generative six-dimensional phase-space reconstruction~\cite{roussel2024gpsr} rather than from single raw diagnostics. The focal position is treated as a noisy condition. Earlier architecture comparisons used flat beam noise with exact $\xof$ and are therefore reported only as auxiliary results.

\section{Methods}\label{sec:methods}
We use a mixture density network because the inverse map is non-unique. An MDN outputs the parameters of a Gaussian mixture, so the probability mass on each solution branch is a number the network predicts and can be read off directly~\cite{bishop1994mdn}. For low-dimensional, strongly multimodal problems, this explicit form has been shown to need less data than implicit generative models such as normalising flows and diffusion models~\cite{guilhoto2026mdnsciml}. The same case has been made in degenerate photonic inverse design, where the predicted mixture is read as a set of competing designs~\cite{unni2020dcmdn}.

\begin{figure*}[htbp]
  \centering
  \includegraphics[width=\textwidth]{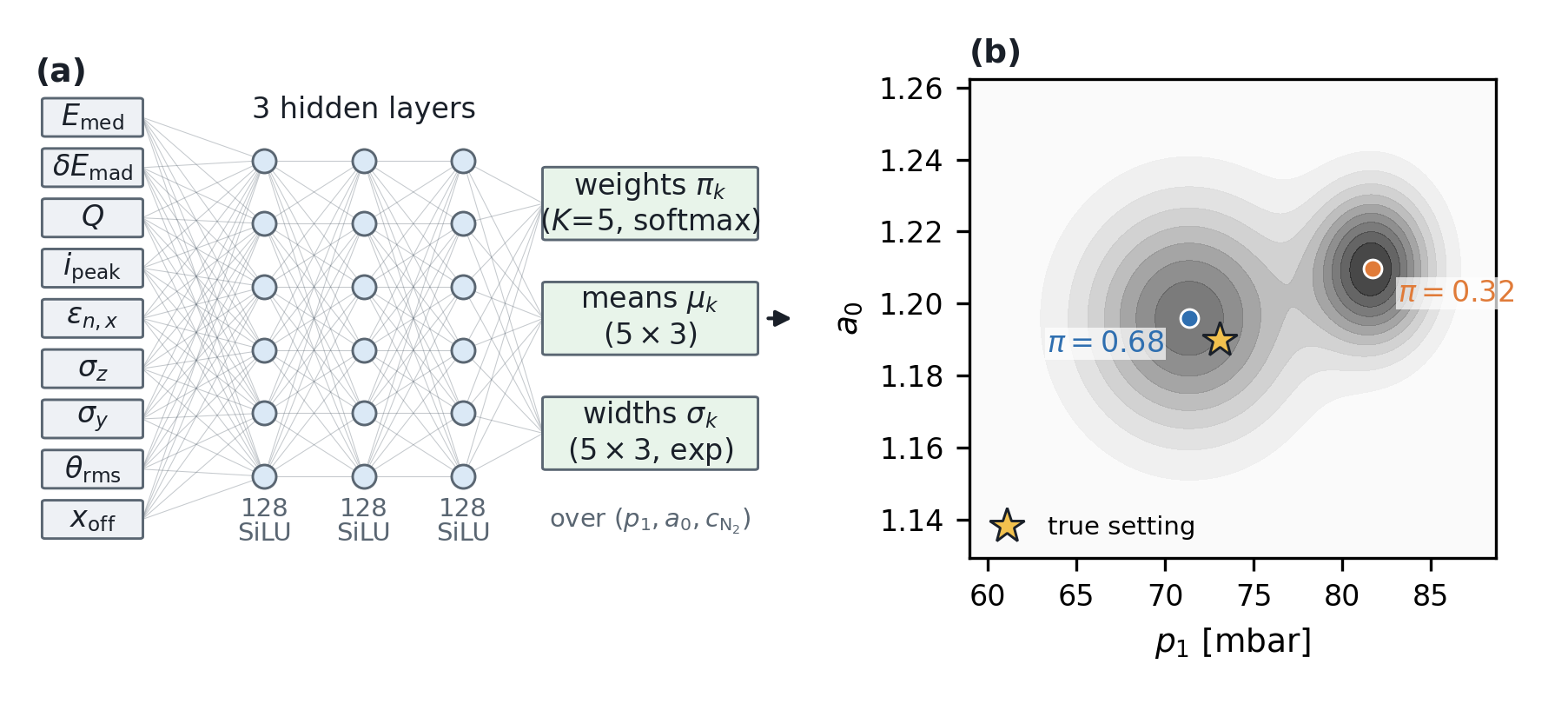}
  \caption{Structure and output of the mixture density network. (a) The nine inputs (eight noisy beam features and the set point $\xof$) pass through three hidden layers of 128 units with SiLU activations. The output layer does not return one value per parameter. It returns the weights $\pi_k$ (softmax), means $\mu_k$ and widths $\sigma_k$ (exponential) of $K=5$ Gaussian components over $(\pone,\azero,\cNtwo)$. (b) The resulting distribution for one test shot, projected on the $(\pone,\azero)$ plane. Each grey peak is one Gaussian component, and the two significant components are marked with their weights $\pi_k$. The star marks the true setting of the presented prediction.}
  \label{fig:mdn}
\end{figure*}

For a given measured beam state \(\mathbf{x}\), the MDN returns a full distribution over \((\pone,\azero,\cNtwo)\). This distribution is the posterior \(p(\pone,\azero,\cNtwo\,|\,\mathbf{x})\), which describes the simulation inputs compatible with the observation, in the sense of~\cite{dopp2023data}. The network provides a Gaussian-mixture approximation to it for any measured beam within the scan range. The prior is the distribution of the training inputs. The scan is uniform, but only the shots with injection are kept, so the prior is uniform over that part of the scan (figure~\ref{fig:setup}(b)). The network has three hidden layers of 128 units with SiLU activations. Its output describes a Gaussian mixture with $K=5$ components over the standardised targets $(\pone, \azero, \cNtwo)$ through the weights $\pi_k$, mean vectors $\mu_k$ and diagonal standard deviations $\sigma_k$. The number of components is set by the sweep of appendix~\ref{app:architectures}. The network is trained by minimising the mixture negative log-likelihood (NLL). Figure~\ref{fig:mdn} illustrates what a single mixture density network returns for one measured beam.

Five MDNs are trained independently and combined into a deep ensemble~\cite{lakshminarayanan2017ensembles}. The ensemble point prediction is the average of the members' posterior means. The predictive variance is the mean of the members' variances plus the variance of the members' means. The pooled-ensemble parameter oracle (defined below) uses all 25 candidate components. The component-level degeneracy and referee analyses instead use the five components of member 2, which has the lowest validation NLL. This avoids treating components from independently trained members as distinct physical branches.

Each member is trained on the injected uniform-scan shots. At each epoch, fresh per-channel noise (see Table~\ref{tab:noise}) is sampled and applied to the training features. This augments the data, and it makes the noise on the training inputs statistically identical to the noise assumed at test time. The full training configuration is given in appendix \ref{app:training}.

Point-prediction accuracy is reported as the coefficient of determination $\Rsq$ of the posterior mean, together with the mean absolute error (MAE) and the absolute percentage error. In addition, we define a \emph{pooled-ensemble parameter oracle}. Each mixture component in each ensemble member proposes a candidate parameter set, and for each shot, the oracle selects the candidate closest to the truth. The $\Rsq$ of these oracle picks quantifies how much accuracy is already contained in the candidate pool, before any selection rule is applied.

Calibration is assessed by the empirical coverage of the central 68\% and 95\% predictive intervals (cov68/cov95). We also recalibrate the intervals with split conformal prediction~\cite{angelopoulos2023conformal}. The test set is split into two. In the first half, we fit the single factor by which the predicted standard deviation must be scaled so that the interval contains the truth with the required fraction of shots, and we apply that factor to the second half. The fitted factor is below the Gaussian value, so the raw intervals are wider than necessary, and the rescaled ones maintain coverage near the nominal level. The guarantee is distribution-free but assumes calibration and test shots are exchangeable (appendix~\ref{app:conformal}).

The choice of the MDN ensemble over alternative posterior architectures and simpler baselines is justified by a comparison in section~\ref{sec:accuracy} (table~\ref{tab:baselines}) and appendix~\ref{app:architectures}.

\begin{figure*}[h]
  \centering
  \includegraphics[width=\textwidth]{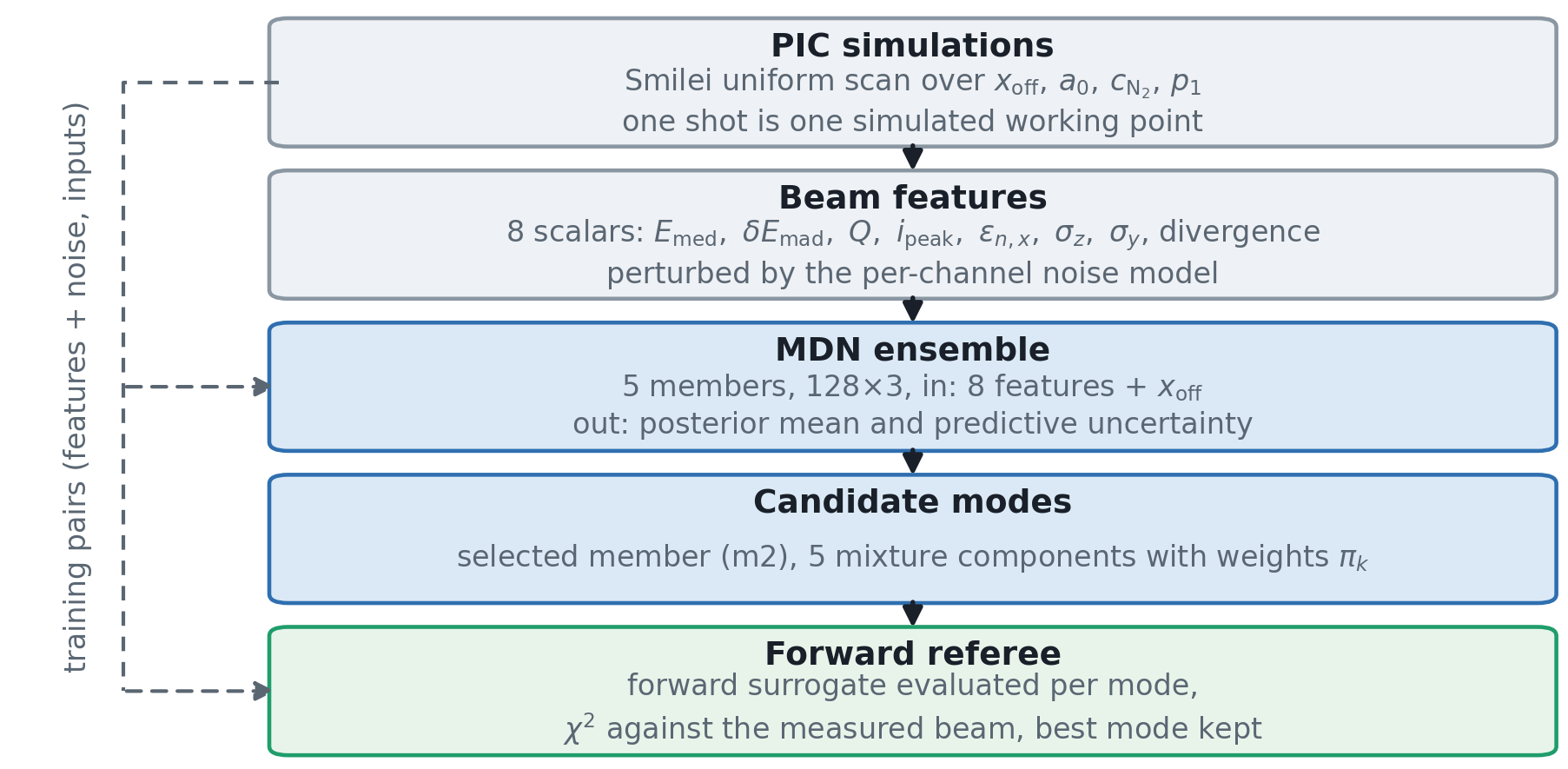}
  \caption{The full sequence overview. Particle-in-cell simulations provide pairs of input parameters and beam features. The eight features are perturbed with the per-channel noise model of table~\ref{tab:noise}. An ensemble of five mixture density networks maps the features and the known $\xof$ to a multimodal posterior over $(\pone, \azero, \cNtwo)$, and a forward surrogate referees the candidate modes against the measured beam. Dashed arrows mark training.}
  \label{fig:sequence}
\end{figure*}

To select a mode in section~\ref{sec:referee}, we also require a fast forward map from the four simulation inputs to the eight beam features. We use a five-member ensemble of heteroscedastic Gaussian MLPs, each with three hidden layers of 256 units, trained on the same uniform-scan shots. At every epoch, we re-sample the per-channel noise of table~\ref{tab:noise} on the beam-feature targets while the input parameters themselves are left clean, the mirror image of the inverse-model training, so that the forward and inverse models assume the same error budget. On held-out data, the ensemble reaches a mean $\Rsq$ of 0.96 across the beam features. Figure~\ref{fig:sequence} summarises the full sequence of models, from the PIC simulation data to the final mode-selection step. 

The inverse model and referee receive the same noisy beam observation. We use each significant component mean as a candidate solution and refer to it as a candidate mode below. For each candidate, the referee score compares its forward prediction with the noisy observation. The $\chi^2$ variance includes the channel-specific measurement variance, the forward-model residual variance and the propagated uncertainty in $\xof$. The score is $\chi^2 = \sum_j (F_j - x_j)^2 / \sigma_j^2$ over the eight beam features, with $F_j$ the forward prediction and $x_j$ the observation. The referee sees only the noisy observation. The clean beam is used afterwards to judge its choice. A candidate is significant when $\pi_k>0.05$ and beam-valid when $\chi^2_8\leq15.51$ against the corresponding clean simulated beam.

\section{Results: accuracy and calibration}\label{sec:accuracy}
\begin{figure*}[h]
  \centering
  \includegraphics[width=\textwidth]{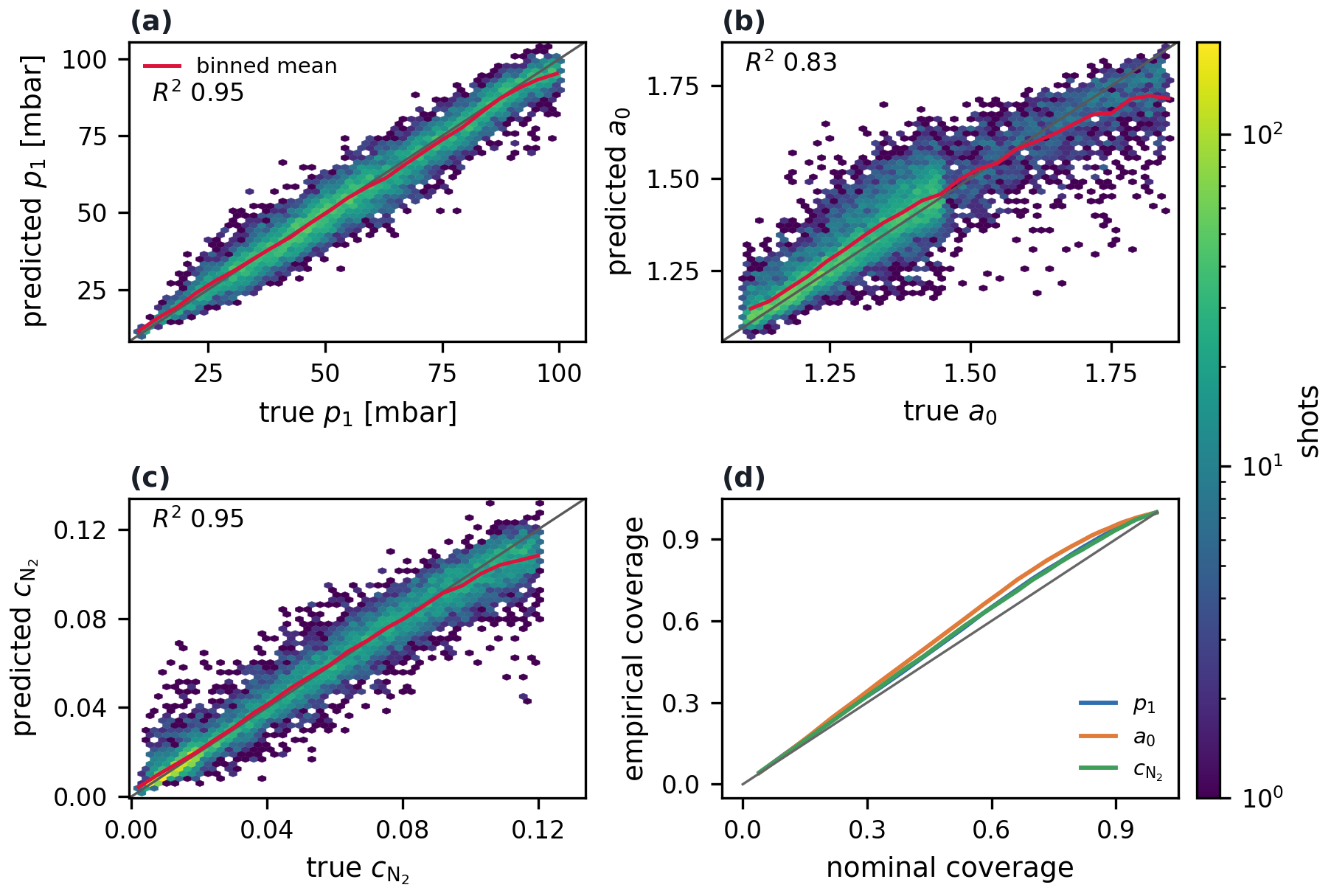}
  \caption{(a)-(c) Posterior-mean predictions versus ground truth for $\pone$, $\azero$ and $\cNtwo$ under nominal per-channel noise (test set, $n=9846$). (d) Empirical coverage of the central predictive intervals against their nominal Gaussian coverage. The diagonal marks perfect calibration.}
  \label{fig:pred_calibration}
\end{figure*}

Under the per-channel noise model, the posterior mean of the MDN ensemble reaches a mean $\Rsq$ of 0.91 on the test set. The normalised vector potential $\azero$ has the lowest $\Rsq$, as shown in table~\ref{tab:baselines}.
It is, nevertheless, the most accurate target in relative terms because its range in the dataset is narrow. Figure~\ref{fig:pred_calibration} shows the predictions against the ground truth. With all noise removed, the same model reaches a mean $\Rsq$ of 0.938. This improvement quantifies the contribution of the assumed diagnostic noise. The residual error contains contributions from inverse ambiguity and finite model and data accuracy (section~\ref{sec:degeneracy}).

\begin{table*}[h]
  \centering
  \begin{tabular}{lccccc}
    \hline
    Method & $\Rsq\,(\pone)$ & $\Rsq\,(\azero)$ & $\Rsq\,(\cNtwo)$ & mean $\Rsq$ & cov68 / cov95 \\
    \hline
    MDN ensemble (per-channel noise) & 0.954 & 0.829 & 0.948 & 0.911 & 0.75 / 0.98 \\
    \hline
    MDN ensemble (flat $20\%$) & 0.911 & 0.693 & 0.889 & 0.831 & 0.72 / 0.97 \\
    MC-dropout                 & 0.950 & 0.814 & 0.942 & 0.902 & 0.85 / 0.99 \\
    Deep ensemble              & 0.954 & 0.839 & 0.947 & 0.913 & 0.67 / 0.95 \\
    cINN                       & 0.953 & 0.833 & 0.948 & 0.911 & 0.69 / 0.95 \\
    cVAE                       & 0.873 & 0.561 & 0.846 & 0.760 & 0.66 / 0.96 \\
    kNN ($k{=}15$)             & 0.848 & 0.459 & 0.851 & 0.719 & -- \\
    Linear regression          & 0.813 & $-0.385$ & 0.785 & 0.404 & -- \\
    \hline
  \end{tabular}
  \caption{Per-target accuracy and calibration. The upper block is the proposed  MDN ensemble under the realistic per-channel noise of table~\ref{tab:noise}. The first row of the lower block is an MDN ensemble trained and evaluated with flat $20\%$ noise and the same $\xof$ uncertainty. The remaining rows are alternative architectures and simple baselines retrained under the same per-channel protocol and scored with a common sample-based evaluator (400 posterior samples). Under that evaluator, the proposed MDN achieves a mean $\Rsq$ of 0.910 with coverage of 0.70 /0.95. Coverage values are means over the three targets.}
\label{tab:baselines}
\end{table*}

An identical MDN ensemble trained and evaluated with flat noise is clearly less accurate, with $\azero$ the most affected target (table~\ref{tab:baselines}, second row). Part of this gap comes from the harder flat test noise. When both models are evaluated under the per-channel noise, training on that profile improves the mean $\Rsq$ by $0.025$. Making $\xof$ exact has a negligible effect for either model, so the difference comes from the beam-channel noise profile.

The remaining rows of table~\ref{tab:baselines} compare the MDN ensemble with four alternative posterior models and two simple baselines. All are retrained under the same per-channel protocol. The alternatives are MC-dropout, a deep ensemble of heteroscedastic Gaussian MLPs, a conditional normalising flow (cINN) and a conditional variational autoencoder (cVAE). The baselines are $k$-nearest neighbours and linear regression. The MDN, the deep ensemble and the cINN agree within $0.003$ in mean $\Rsq$, but the unimodal deep ensemble cannot place probability mass on a second solution. The cINN has the best negative log-likelihood, $0.72$ below the MDN, at the same accuracy (table~\ref{tab:arch}). We keep the MDN. Its components are explicit, and it trains ten times faster. MC-dropout comes close to their point accuracy but over-covers. The simple baselines fall far behind, confirming that a non-linear model is warranted. The full comparison is given in appendix~\ref{app:architectures}.

The Gaussian intervals built from the predictive variance are slightly too wide, so they over-cover, as shown in figure~\ref{fig:pred_calibration}(d) and table~\ref{tab:baselines}. Intervals sampled from the mixture itself are close to nominal (table~\ref{tab:baselines} caption, appendix~\ref{app:conformal}). Their relative sizes are nevertheless meaningful. Shots with a larger predicted standard deviation tend to have larger errors.

For the present ensemble and per-channel measurement model, the median variance decomposition is dominated by the aleatoric term, which includes measurement noise and inverse-map degeneracy. The epistemic term, the one that more or better-targeted training data would reduce, is the smaller of the two. The margin for improvement is therefore mainly on the measurement side. A new diagnostic channel could break the degeneracy, or the existing channels could be made more precise.

We also tested the sensitivity of the accuracy to each noise channel separately. Increasing the noise on the divergence, bunch length, or charge by a factor of 4 reduces the mean \(\Rsq\) by $0.08$-$0.11$. In contrast, improving any single channel beyond the nominal level increases the mean \(\Rsq\) by at most $+0.007$  (appendix~\ref{app:noise}). This suggests that, in the nominal setting, the remaining error is not controlled by a single easily improved diagnostic channel.

\section{Degeneracy}\label{sec:degeneracy}
The posterior offers several candidate solutions for each shot. The choice among them affects the accuracy. Choosing the candidate whose forward beam best matches the measurement raises the fraction of shots on which the selected candidate is the one closest to the truth from $0.548$ to $0.617$ (section~\ref{sec:referee}, table~\ref{tab:referee}). Part of the point-estimate error is therefore made at the selection step. Candidate-set failures also remain possible (section~\ref{sec:referee}). This section characterises the competing candidates.

\subsection{Dual posterior modes}\label{sec:deg:branches}
\begin{figure*}[h]
  \centering
  \includegraphics[width=\textwidth]{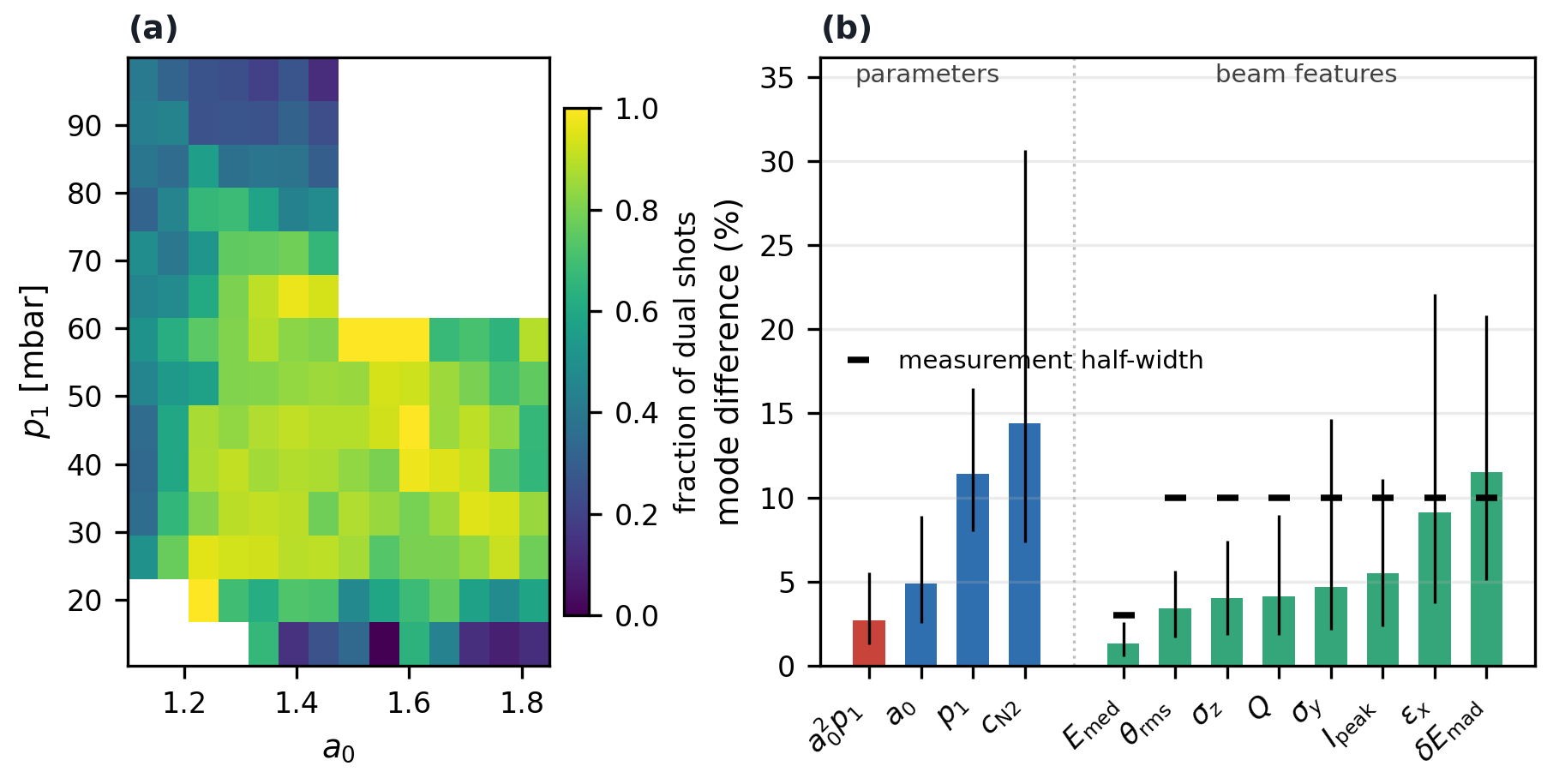}
  \caption{(a) Fraction of dual shots ($\pi_2 > 0.20$) across the $(\azero, \pone)$ plane of the true test parameters. White bins contain no injected shots. (b) Median relative difference between the two leading posterior modes over the $6411$ dual shots, with error bars marking the 25th and 75th percentiles. Red: in the parameters, the two solutions differ by $5$-$14\%$ each, while the self-focusing strength combination $\azero^2\pone$ is conserved to $2.7\%$. Black: in the forward-propagated beam features, the separation between the two solutions stays below the assumed measurement half-widths of table~\ref{tab:noise} for seven of the eight features. The energy spread $\dEmad$ is the only marginal exception ($11.5\%$ against a $10\%$ half-width).}
  \label{fig:degeneracy_atlas}
\end{figure*}

We call a test shot \emph{dual} when the second mixture component of the MDN posterior has weight $\pi_2 > 0.20$, and we call the two leading components its branches. The two leading components are separate peaks. For $92\%$ of the dual shots, the density drops between the two means. For the rest, the two components overlap. Under the per-channel noise model, $65\%$ of the test shots meet this parameter criterion for the validation-selected member. The fraction depends on the threshold. At $0.15$ it is $75\%$ and at $0.25$ it is $46\%$. The beam-space count of section~\ref{sec:referee}, which uses no weight threshold, gives $66\%$. The exact rate varies among ensemble members, but the duality persists when all noise is removed. The degeneracy is therefore not an artefact of the measurement noise. The branch weights are close to calibrated. A component with weight $\pi_k$ is the one closest to the truth in about a fraction $\pi_k$ of the shots. Figure~\ref{fig:degeneracy_atlas}(a) shows that dual shots are concentrated towards the upper end of the scanned laser-strength range and at low to intermediate pressure.

Between the top two modes, the individual parameters differ by $5$-$14\%$, while the combination $\azero^2\pone$ is conserved to $2.7\%$ in the median (figure~\ref{fig:degeneracy_atlas}(b)). This combination sets the strength of relativistic self-focusing. The critical power for self-focusing is $P_c \simeq 17\,(\lambda_p/\lambda)^2$~GW~\cite{esarey2009physics}. With the focal spot and pulse duration fixed, the laser power scales as $P \propto \azero^2$. At fixed nitrogen concentration, the plasma density scales as $n_e \propto \pone$. Thus, $P/P_c \propto \azero^2\pone$. The two modes also differ in $\cNtwo$, which changes the density through the nitrogen electrons. This change partly compensates the change in $\azero^2\pone$, so that $\azero^2 n_e$ is conserved slightly better than $\azero^2\pone$ ($2.3\%$ against $2.7\%$ in the median, counting nitrogen ionised to N$^{5+}$).

Beams with the same self-focusing strength but different $(\azero,\pone)$ splits produce nearly the same measured features. We show this with the forward surrogate, so the statement is only as accurate as the surrogate (figure~\ref{fig:degeneracy_atlas}(b)). For seven of the eight features, the median separation between the two modes is below the assumed measurement half-width of table~\ref{tab:noise}. The energy spread $\dEmad$ is the only marginal exception, with a median separation of $11.5\%$ against a $10\%$ half-width.

The degeneracy is approximate. A Jacobian analysis of the forward surrogate at fixed $\xof$, using the median singular values of $\partial(\ln \mathrm{beam})/\partial(\ln \pone, \ln \azero, \ln \cNtwo)$ over 500 test settings, gives a condition number of about 18. No direction of parameter space is strictly unobservable. The beam responds to the weakest direction eighteen times more weakly than to the best-measured one. The assumed measurement noise hides this weak response. Along this weakest direction $\azero^2\pone$ stays nearly constant. Most of the change is in $\cNtwo$, consistent with the $14\%$ median $\cNtwo$ difference between the two modes (figure~\ref{fig:degeneracy_atlas}(b)). With $\xof$ as a fourth input, the condition number rises to about 115. The weakest direction is dominated by $\xof$. This is why $\xof$ is supplied as a known input (section~\ref{sec:setup:features}). Two observations support this picture. $\azero$ remains partially recoverable ($\Rsq = 0.83$, well above zero), and the two modes stay marginally separated in $\dEmad$. A more precise measurement of the separating features would recover the lost direction (section~\ref{sec:discussion}).

\subsection{The continuous same-beam valley}\label{sec:deg:valley}

The two posterior modes of section~\ref{sec:deg:branches} could in principle be an artefact of the mixture representation. This section shows that there are two points on a continuous valley in the forward map itself, and measures the direction and strength of this valley using methods that depend as little as possible on the learned models. The procedure is as follows. We take a target $\thetastar=(\pone,\azero,\xof,\cNtwo)$ and compute its beam $\bstar=\mathrm{forward}(\thetastar)$. We then sweep the $(\pone,\azero)$ plane at fixed $\xof$ and $\cNtwo$ and score each point by the reduced $\chi^2$ between its beam and $\bstar$, in units of the per-channel measurement noise. Points with reduced $\chi^2 \le 1$ produce beams that are indistinguishable from the target under realistic diagnostics, so we call this set the same-beam region.

\begin{figure*}
  \centering
  \includegraphics[width=\textwidth]{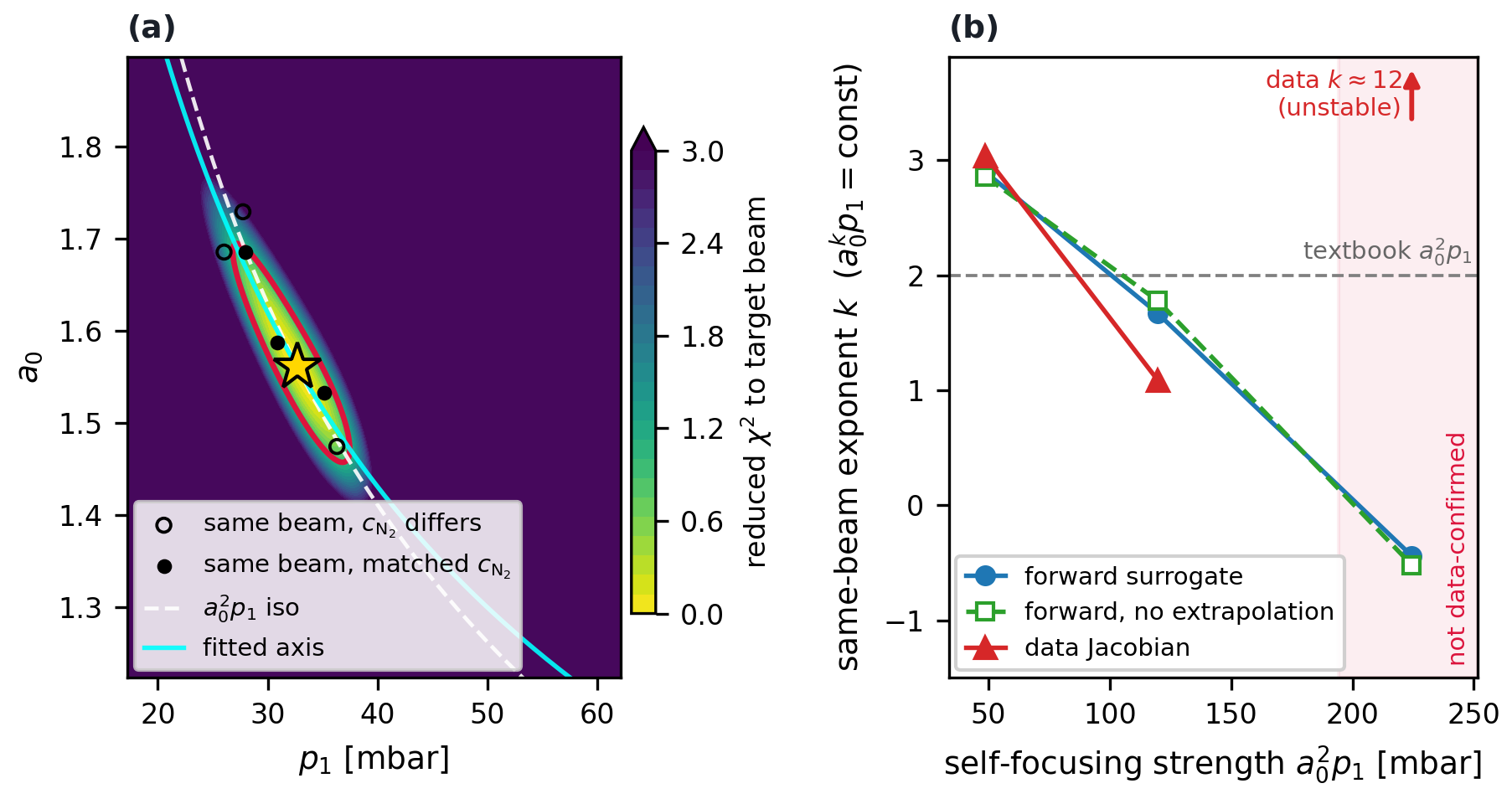}
  \caption{The continuous same-beam degeneracy. (a) Reduced $\chi^2$ between the forward-model beam and a data-confirmed target beam ($\azero^2\pone \approx 80$, star) across the $(\pone,\azero)$ plane at fixed $\xof$ and $\cNtwo$. The crimson contour marks reduced $\chi^2 = 1$ and encloses the same-beam region; the white dashed line is the $\azero^2\pone$ iso-curve, and the cyan line is the fitted valley axis. Black points are the seven real simulations (of 3758) that reproduce the target beam within this budget at $\xof$ matched to $\pm 200~\mu$m, filled where $\cNtwo$ also matches and open where it differs. (b) Fitted valley exponent $k$ of $\azero^k\pone=\mathrm{const}$ against target self-focusing strength $\azero^2\pone$, estimated in three ways: from the forward surrogate, from a clean-trained surrogate with extrapolation forbidden, and from the raw-data local Jacobian. The three estimates agree on $k \approx 2.9$ at low density, above the textbook $\azero^2\pone$ scaling, and the valley weakens towards a nearly unique solution ($k \approx 1.1$ from the data, $1.7$ from the surrogate) at mid density. In the shaded high-strength region, the surrogate extrapolates, and the data Jacobian is unstable, so the valley there is not confirmed by the data.}
  \label{fig:degregion}
\end{figure*}

At the low-density end of the scan, the same-beam region is a long, thin valley (figure~\ref{fig:degregion}(a)). Its direction is estimated in three ways: from the forward surrogate, from a clean-trained surrogate with extrapolation forbidden, and from a local Jacobian built on neighbouring simulations. All three agree on a valley near $\azero^{2.9}\pone=\mathrm{const}$, steeper than the textbook $\azero^{2}\pone$ scaling. The two discrete candidate modes lie close together in this valley, which explains why $\azero^{2}\pone$ is approximately conserved between them.

As the density rises, the valley weakens (figure~\ref{fig:degregion}(b)). Near the median target, the raw-data Jacobian gives an exponent close to one ($1.1$) and a condition number of about three. The surrogate gives an exponent of $1.7$. The beam there is nearly uniquely identifiable. At the upper end of the scanned range of $\azero^2\pone$, the surrogate finds no second valley, and the raw-data Jacobian is unstable there. The data therefore support a single degeneracy, which scales approximately as $\azero^{2.9}\pone$ at low density and approaches a nearly unique solution at higher density. Because the degeneracy is approximate, a second measurement at a different density can break it (section~\ref{sec:twowp}).

\section{Reducing the degeneracy}\label{sec:reducing}
The ambiguity of section~\ref{sec:degeneracy} can be reduced by providing the model more information about each shot, or by testing the candidate solutions against a forward model. This section evaluates both and then presents a surrogate calculation and a paired PIC test of how a second measurement at a known density reduces the ambiguity.

\subsection{Adding the electron spectrum}\label{sec:spectrum}
\begin{figure*}
  \centering
  \includegraphics[width=\textwidth]{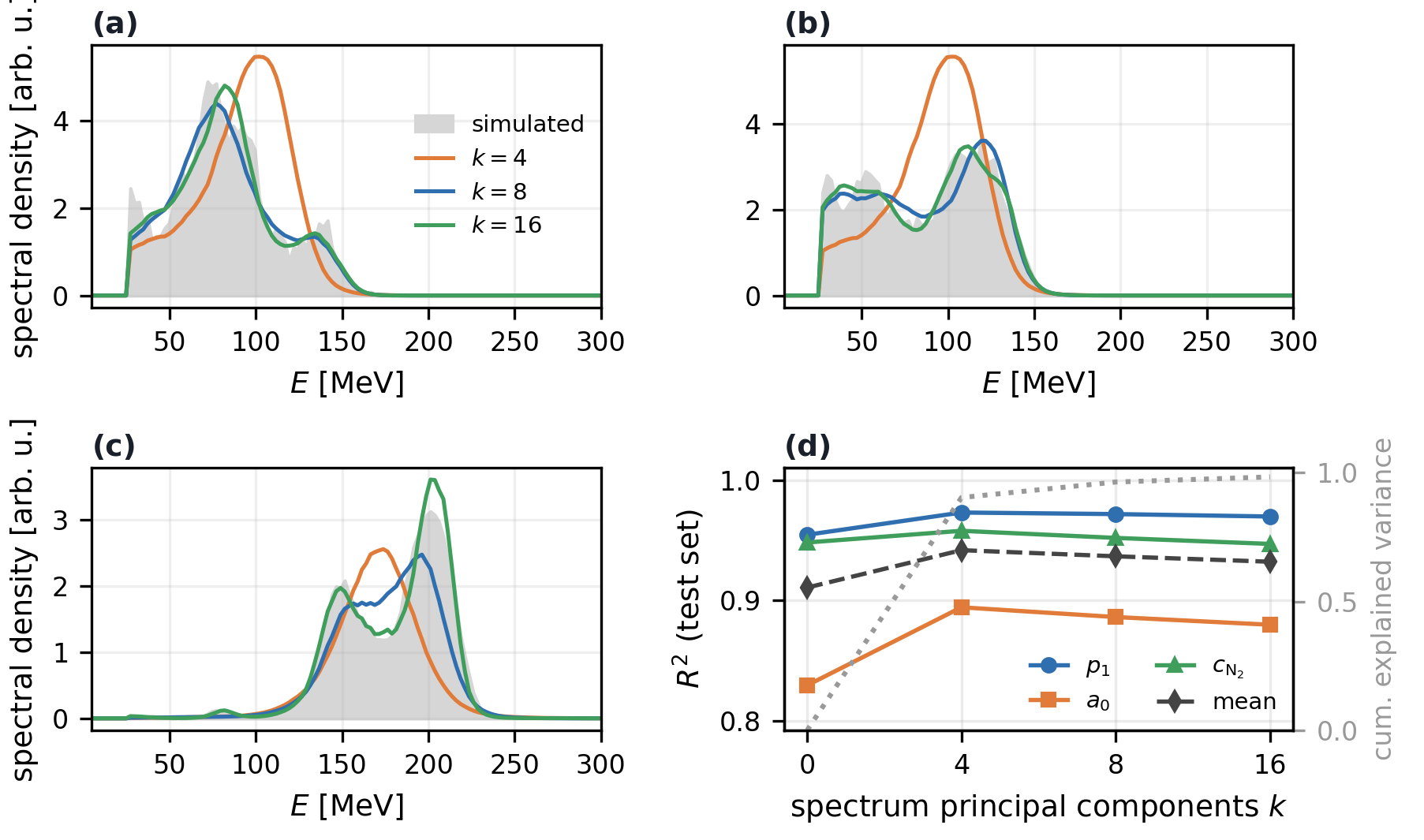}
  \caption{(a)-(c) Three simulated test spectra and their principal component reconstructions with $k=4$, 8 and 16 components. The $k=4$ reconstruction misplaces peaks and merges structure. Its relative $L_1$ error over the full test set is 63\%, against 35\% at $k=8$ and 16\% at $k=16$. (d) Accuracy on the test set ($n=9846$) against the number of retained components $k$. The $k=0$ anchor is the scalar-only model under the same channel noise. The grey dotted line is the cumulative explained variance of the PCA basis. The mean $\Rsq$ peaks at $k=4$ and declines slowly for larger $k$. The components that carry the fine structure do not help the inversion.}
  \label{fig:spectrum}
\end{figure*}

The scalar feature set compresses the energy spectrum into two numbers, $\Emed$ and $\dEmad$. The full spectrum may retain information that this compression destroys. To test this, we re-grid each simulated spectrum onto a common energy axis with 160 bins from 5 to 400~MeV, log-compress it, standardise each bin, and reduce it using principal component analysis fitted to the training set. Eight components capture 96.4\% of the variance (figure~\ref{fig:spectrum}(d)).
The model inputs are the nine noisy scalar features plus the eight leading components of the noisy spectrum. Each bin receives a $\pm 10\%$ perturbation, and the overall amplitude is further perturbed by $\pm 10\%$, re-sampled every epoch. Computing $\Emed$ and $\dEmad$ from the same noised spectrum, instead of noising them separately, changes the mean $\Rsq$ by less than $0.01$ in favour of the spectrum model. The separate noise does not inflate the gain. The training protocol and test set are identical to those used by the proposed model.

Adding the spectrum helps $\azero$ most. Its score rises from $\Rsq=0.83$ to $0.89$, and the mean over the three inferred parameters rises from $0.91$ to $0.94$ with eight components. The accuracy of the closest candidate mode does not change. The spectrum changes the weights of the candidate modes. It does not move the modes.

Figure~\ref{fig:spectrum} shows the reconstructions and the accuracy against the number of retained components. Four components give the best accuracy, although their relative $L_1$ reconstruction error is $63\%$. The optimum is shallow. Eight components are within $0.005$ and sixteen within $0.01$. The gain from the spectrum does not depend on this choice. We did not test it under other noise assumptions. Panels (a) to (c) show that this coarse representation misplaces peaks and merges the double-peaked structure. Retaining the components that restore this structure consistently reduces inversion accuracy across all ensemble members. The higher components carry little variance. Noise can dominate them and spoil the fit. The present experiment does not identify their physical origin. Four components are sufficient for the inversion.

\subsection{Mode selection with a forward surrogate}\label{sec:referee}
The posterior already provides a small set of candidate solutions, so this route needs no additional measurement. We pass each candidate through the forward surrogate, compare its predicted beam with the measured beam using the $\chi^2$ score of section~\ref{sec:methods}, and retain the candidate with the lowest score.

We evaluate the referee on the random-scan test set ($n=9846$) under the per-channel noise model. Each selection rule is judged against two distinct objectives. A mode is \emph{beam-valid} if its forward beam matches the clean target beam within the error budget, $\chi^2_8 \le 15.51$, the 95th percentile of the $\chi^2$ distribution with eight degrees of freedom, one per beam feature. A mode is \emph{branch-correct} if it lies closest to the true parameters among the candidates. The two objectives do not coincide, because a wrong branch can still reproduce the beam almost exactly. Table~\ref{tab:referee} scores every rule against both. The true parameters meet the threshold in $93\%$ of shots, not $95\%$. The missing shots are those where the forward surrogate is least accurate. The error budget assigns the same surrogate error to every shot. Given the error the surrogate ensemble predicts for each shot, the true parameters meet the threshold in $97\%$ of cases. We keep the same error for every shot. It rejects more candidates on those shots, so it does not inflate the degeneracy.

In beam space, $66\%$ of the shots have at least two significant beam-valid modes, close to the $65.1\%$ that are dual under the parameter criterion of section~\ref{sec:deg:branches}. One in ten shots has none. For these shots, the candidate set itself fails, and no selection rule can repair it.

The largest-weight mode is already beam-valid for most shots. The referee raises this fraction further, to $93\%$ on dual shots (Table~\ref{tab:referee}).

\begin{figure*}[h]
  \centering
  \includegraphics[width=\textwidth]{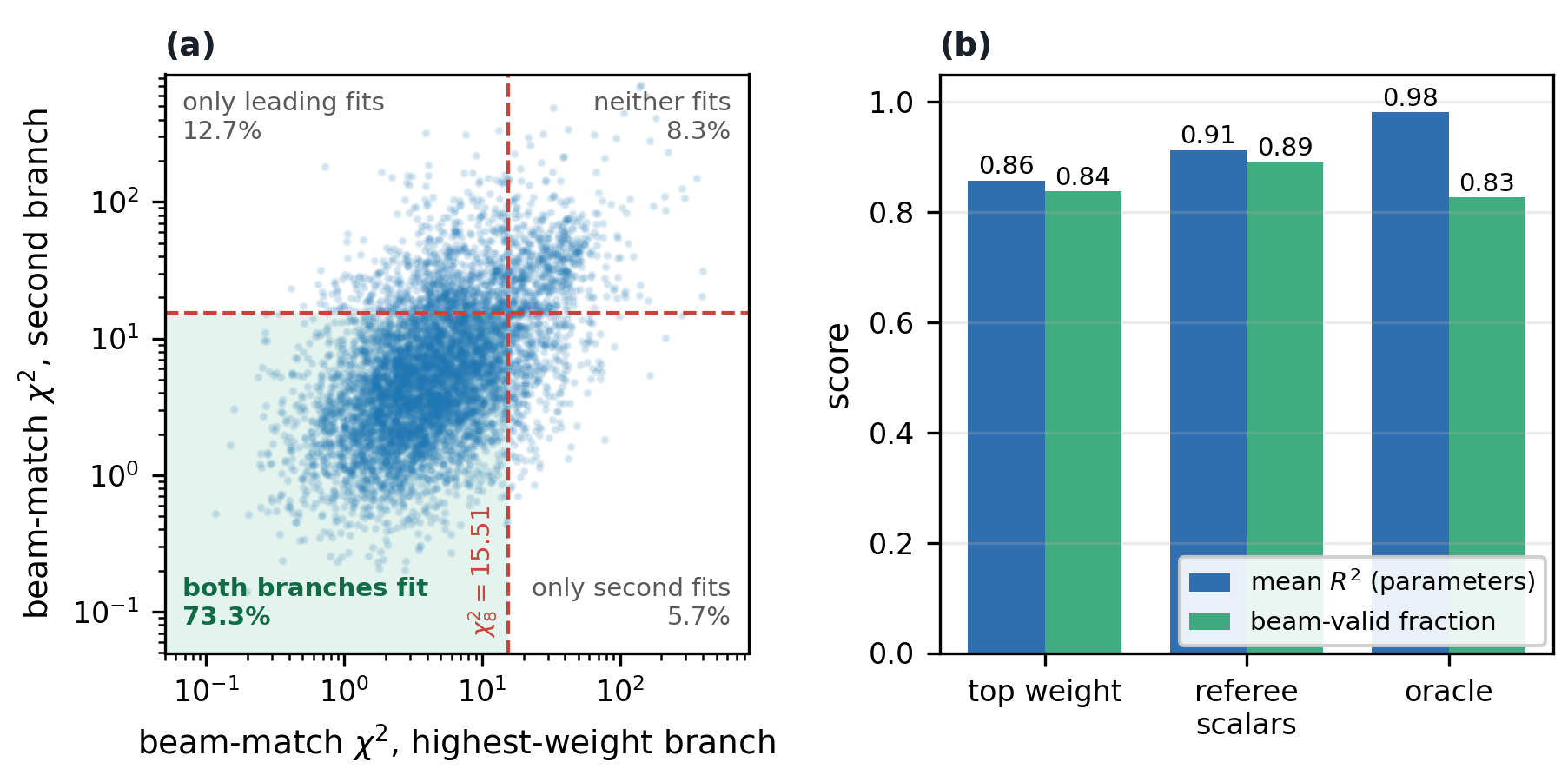}
    \caption{Forward-referee mode selection. (a) Each point is one dual shot under per-channel noise ($n=6411$), placed by the beam-match $\chi^2$ of its highest-weight branch (horizontal) and of its second branch (vertical). The dashed lines mark the validity threshold $\chi^2_8 = 15.51$. For $73.3\%$ of the shots, both branches fall below it (shaded quadrant), so both reproduce the measured beam and the beam cannot separate them. (b) Mean parameter $\Rsq$ of the selected mode and fraction of beam-valid selections for each rule. The referee lifts both scores over the top-weight rule. The parameter-oracle mode, the candidate closest to the truth, is beam-valid on only $82.6\%$ of shots.}
  \label{fig:referee}
\end{figure*}

On dual shots, the referee improves on the top-weight rule but identifies the true branch for barely half of them. Rebuilding both the candidates and the referee inputs from clean features raises this only to $62\%$. The branch ambiguity survives even a noise-free pipeline. As figure~\ref{fig:referee}(a) shows, on $73.3\%$ of the dual shots, both leading branches reproduce the measured beam within the budget, so the eight beam features barely distinguish them.

The parameter-oracle row of Table~\ref{tab:referee} makes a related point. The candidate closest to the truth is not always the one that best reproduces the beam. Parameter accuracy and beam fidelity are different objectives, and a referee tuned to the beam can therefore exceed the parameter oracle under the beam metric.

Auxiliary flat-noise tests suggest the saturation is not specific to the mixture density network. A classifier referee trained directly on the features in the spirit of simulation-based inference~\cite{cranmer2020sbi}, a normalising-flow posterior~\cite{ardizzone2019inverse} and a cascade conditioned on the predicted $\pone$ all leave the branch ambiguity unresolved.

\begin{table*}
  \caption{Referee variants under the per-channel noise model using the five candidates of the validation-selected member. Parameter correct means that the selected candidate is closest to the truth. Branch correct applies the same criterion to dual shots. Beam-valid means that the selected candidate satisfies $\chi^2_8 \le 15.51$ against the clean beam.}
  \label{tab:referee}
  \centering
  \begin{tabular}{lcccc}
    \hline
    Selection rule & Parameter correct & Branch correct & Beam-valid & Beam-valid (dual) \\
    \hline
    Top weight (no referee)      & 0.548 & 0.429 & 0.837 & 0.861 \\
    Referee: scalars only        & 0.617 & 0.541 & \textbf{0.890} & \textbf{0.927} \\
    Parameter oracle             & 1.000 & 1.000 & 0.826 & 0.862 \\
    \hline
  \end{tabular}
\end{table*}

\subsection{A second measurement at a different density}\label{sec:twowp}
The degeneracy of section~\ref{sec:degeneracy} is approximate, and its orientation depends on the plasma density. Thus, a second beam measurement taken after a known density change carries information that the first measurement does not. The forward surrogate predicts the size of the effect. Crossing the consistent regions of two working points separated by a factor of $2.1$ in density narrows the range of $\azero$ from $3$-$7\%$ to $1$-$4\%$ across five targets. Multiple settings sharing $\azero^2\pone$ reach the same self-focusing strength, but the weak residual dependence on $\azero$ and $\pone$ through injection and dephasing shifts with density, so the two valleys cross only near the true setting.

\begin{figure*}
  \centering
  \includegraphics[width=\textwidth]{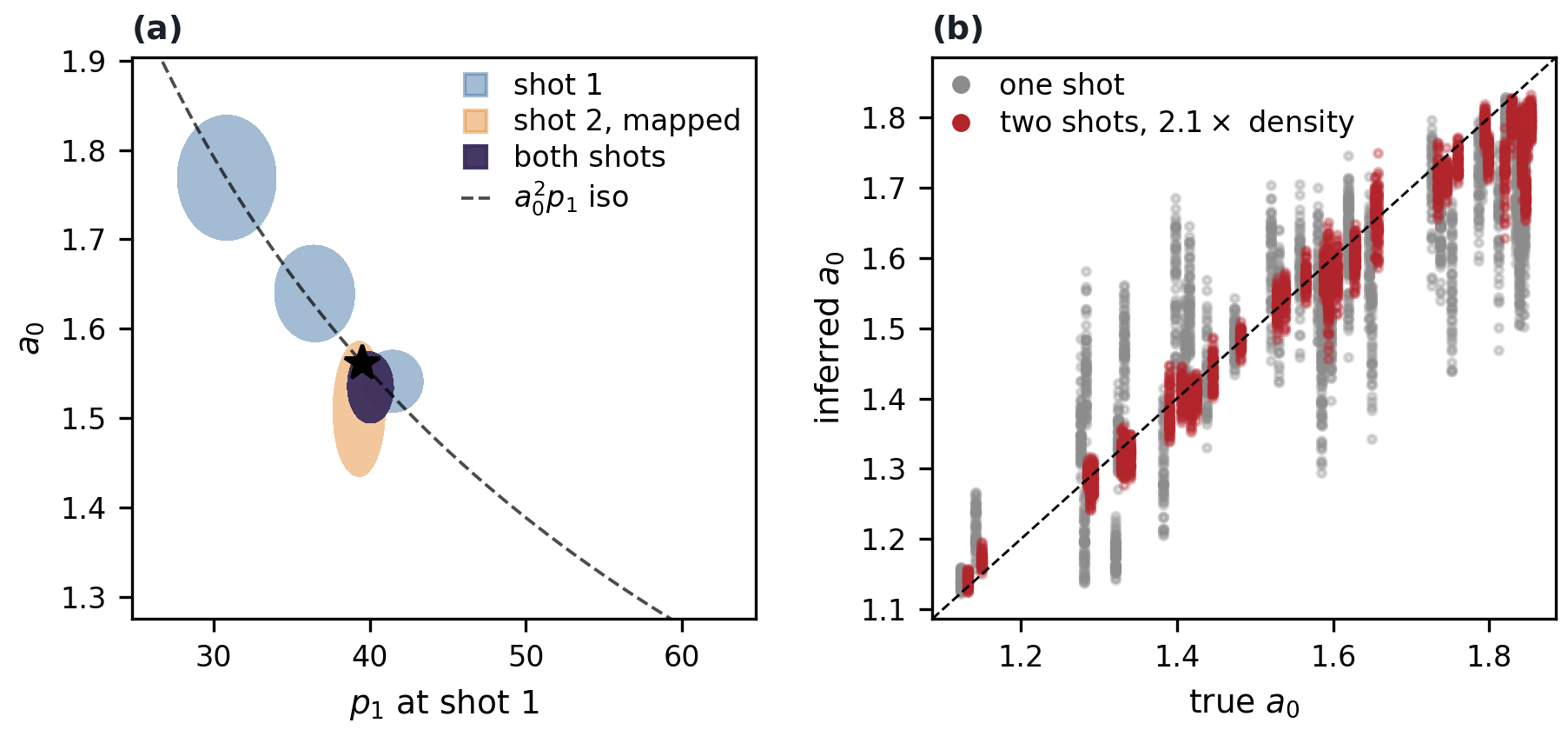}
    \caption{Confirmation of the two-working-point gain on particle-in-cell beams, using the frozen inverse model of section~\ref{sec:methods} with no retraining and no surrogate in the loop ($n=38$ pairs, 64 noise realisations each). (a) The mechanism for the pair with the median gain in $\azero$. The posterior of shot 1 is blue, the posterior of shot 2 mapped into the coordinates of shot 1 is orange, and their product is dark. The star marks the true setting. The two posteriors are elongated along differently oriented valleys, so their product survives only where they cross. (b) Inferred against true $\azero$ for every pair and every noise realisation, with one shot in grey and with two shots at $2.1$ times the density in red. The dashed line marks exact recovery. The second shot reduces the RMSE of the posterior mean by factors of $4.05$, $2.59$ and $2.95$ for $\pone$, $\azero$ and $\cNtwo$, against $1.12$, $1.09$ and $1.12$ for a repeat shot at the same density, which supplies the same averaging without a density change.}
  \label{fig:pic_twowp}
\end{figure*}

We tested this in a dedicated paired particle-in-cell campaign, applying the ensemble of section~\ref{sec:methods} unchanged to the simulated beam features, with no retraining and no surrogate in the loop. The campaign is described in appendix~\ref{app:pic}. Against a single shot at the lower density, the pair reduces the RMSE of the posterior mean by factors of $2.6$ to $4.0$ across the three parameters, and a bootstrap over the 38 pairs keeps all three above $2$. A repeat shot at the same density stays below $1.15$ (figure~\ref{fig:pic_twowp}). The mass carried by the second posterior mode falls by one to two orders of magnitude for $\pone$ and $\azero$, where a repeat shot barely changes it.

Moving a single shot to the higher density already reduces the error. Pairing it with the lower-density shot further reduces it. For $\azero$ and $\cNtwo$, the pairing is the larger step. For $\pone$ the two steps are about equal, and its total gain is the largest of the three. All pairing factors lie well above the repeat-shot control (table~\ref{tab:picpairs}).

The 38 pairs lie at the low-pressure end of the scan, with the lower-density shot between $12$ and $48$ mbar, where section~\ref{sec:deg:valley} finds the valley to be longest. Thus, the test probes the regime in which a second working point is expected to help most and does not test the high-density regime. With 38 pairs, gain factors below about $1.5$ are not resolved. The three pairing factors lie above this limit. The density step for $\azero$ and $\cNtwo$ lies below it. Appendix~\ref{app:pic} reports the empirical coverage on the 38 pairs and the conditional-independence assumption behind the posterior product. Both shots come from the simulator that produced the training library. The test validates the inference method, but not the simulator.

\section{Discussion}\label{sec:discussion}

We have trained an ensemble of mixture density networks to infer the LPI input parameters from a measured electron beam under a per-channel noise model. Part of the remaining point-estimate error comes from candidate selection. This is visible in the referee result of section~\ref{sec:referee}, where choosing among the candidates raises the fraction of correctly selected candidates from $0.548$ to $0.617$. The candidate list does not contain a correct solution for every shot, as one shot in ten has no candidate that reproduces the beam.

The remaining selection error is consistent with the physical degeneracy of section~\ref{sec:degeneracy}. Posterior models built on quite different assumptions reach essentially the same accuracy (appendix~\ref{app:architectures}). Even with clean features and clean candidates, the true branch of a dual shot is identified only modestly more often than under noise (section~\ref{sec:referee}).

These results suggest that the diagnostics limit the accuracy more than the model does. Further modelling or targeted training data could still help.

The second working point of section~\ref{sec:twowp} is confirmed on 38 paired PIC simulations. It still needs an experimental test. A more accurate energy spread measurement could also help, although the case for it is weaker than the raw ranking suggests. $\dEmad$ shows the largest median difference between the branch predictions, but the forward surrogate predicts this feature with a comparable error, so the apparent separation cannot be distinguished from model error. Judged against the surrogate error, the charge is the most reliable discriminant, although its branch separation remains below the assumed measurement accuracy. A meaningful precision sweep would first require a forward model whose $\dEmad$ error is well below the branch separation. In practice, the accuracy of a measured $\dEmad$ is set by the stability of the electron beamline transport as much as by the spectrometer itself.

The single-shot model can also serve as a virtual diagnostic during operation. In a simulated drift run (appendix~\ref{app:drift}), it recovers the imposed $\azero$ trend and correctly reports a constant $\cNtwo$. The recovered $\pone$ shows a small spurious trend. Monitoring $\azero^2\pone$ instead tracks the true change to within one percentage point.

The same combination links the inverse model to the earlier PALLAS design studies. In those studies, injection required self-focusing. Over the scanned range of $\azero$, nitrogen is ionised to N$^{5+}$ and N$^{6+}$ only after the laser self-focuses. The injection boundary follows the critical-pressure curve $p_c \propto \azero^{-2}$~\cite{kane_direct_model,drobniak2023random}. This curve is a line of constant $\azero^2\pone$. The same quantity sets the degeneracy of the injected beams.

Along the valley, a change of $\azero$ compensated by $\pone$ leaves the beam unchanged within the assumed measurement accuracy. This makes the degeneracy useful for LWFA operation. The same-beam region of figure~\ref{fig:degregion}(a) is a tolerance map for $\azero$ and $\pone$. The tolerance is the largest at the low-density end of the scan, where the valley is the longest. This tolerance applies only along the valley. The beam stays sensitive to $\xof$, $\cNtwo$ and the direction across the valley.

The feature set includes the bunch length $\sigma_z$, which is not measured at PALLAS. Under the auxiliary flat-noise protocol, removing it lowers the mean $\Rsq$ by $0.024$ for the base model and by $0.010$ when the spectrum is added. The conclusions of sections~\ref{sec:degeneracy} and~\ref{sec:reducing} do not depend on this feature. In addition, all models are trained and evaluated inside the scanned parameter ranges of section~\ref{sec:setup:sims}. Predictions for beams produced outside this operating window are extrapolations and carry no guarantee of accuracy. All other laser and gas parameters are fixed in the training data. In the experiment, they vary from shot to shot. The model does not know this. The results do not depend on the shape of the noise. Gaussian noise with the same variance gives the same accuracy and the same dual fraction. The same holds when the channels of one instrument share a single error. A fixed calibration error at the edge of the budget costs up to $0.03$ in mean $\Rsq$ and shifts the parameters by up to $4\%$. Finally, all beams in this study are simulated, and calibration offsets or correlated measurement errors are not included in the noise model.

\section*{Acknowledgments}
This work was granted access to the HPC resources of TGCC Irene Joliot-Curie under the allocations 2021-A0110510062 and 2022-A0130510062, and to the HPC resources of CINES Adastra under the allocation 2025-A0190510062, made by GENCI for the project Virtual Laplace. The work was supported by CNRS IRL DMlab and PACRI – European Union Grant Agreement n. 101188004.

\section*{Data availability statement}
The data that support the findings of this study are openly available at the following URL/DOI: \url{https://doi.org/10.5281/zenodo.23156728}~\cite{lpi_zenodo_data}. This record contains the training and test datasets (SET1, SET2) and the paired particle-in-cell runs of appendix~\ref{app:pic}. The code, trained models and notebooks that reproduce every figure and table of this article are available at \url{https://gitlab.in2p3.fr/lpa-pic-simulations-data/inverse-modelling-lpa}~\cite{lpiinverse_data}. Raw PIC simulation data are available from the corresponding author upon reasonable request.

\appendix

\section{Training configuration}\label{app:training}
The posterior is a five-component Gaussian mixture over the three targets, produced by a three-layer network of width 128; the log-variances are clamped to $[-10,5]$. A single random $15\%$ of the training set, drawn once and shared by all five members, is held out for validation. We train with the Adam optimiser at a learning rate of $2\times10^{-3}$, with cosine annealing over 400 epochs, weight decay of $10^{-5}$, and a batch size of 256, on the plain mixture negative log-likelihood. Gradient norms are clipped at 5, which suppresses the loss spikes that the likelihood produces on shots with small predicted variance. The per-channel measurement noise is re-drawn for every training batch at every epoch (section~\ref{sec:setup:noise}). The validation features are once-noised with a fixed seed, so that changes in the early-stopping signal come solely from the weights. Training stops early when the validation negative log-likelihood has not improved by more than $10^{-4}$ for 80 epochs, and the weights of the best validation epoch are kept. The five members differ only in their initialisation and noise seed (100-104). The validation split is common to all of them, so their spread reflects only initialisation and noise variability.

\section{Conformal calibration of the predictive intervals}\label{app:conformal}
We calibrate the $90\%$ predictive intervals with split conformal prediction under the per-channel noise model and the assumption of exchangeability. The test set is reproducibly divided into two halves of 4923 shots each, one for calibration and one for evaluation. The raw interval is the Gaussian one, $\pm1.645$ times the total predictive standard deviation obtained by combining the component variances and the spread of the five members. Table~\ref{tab:conformal} shows that it over-covers all three parameters, so the fitted scale factors fall below $1.645$ and calibration narrows the intervals while holding coverage near nominal. The over-coverage comes from the Gaussian summary. The conformal guarantee covers this central interval only. On a dual shot, the interval spans both modes and the gap between them. It says nothing about the shape of the posterior. The NLL of appendix~\ref{app:architectures} does. A single Gaussian carrying the total variance of a multimodal mixture is wider than the $90\%$ region of the mixture itself, and the same effect separates the analytic from the sample-based coverages of table~\ref{tab:baselines}.

\begin{table}[h]
  \caption{Split conformal calibration of the $90\%$ predictive intervals of the proposed model.}
  \label{tab:conformal}
  \centering
  \small
  \begin{tabular}{lccc}
    \hline
    & $\pone$ & $\azero$ & $\cNtwo$ \\
    \hline
    raw coverage          & 0.945 & 0.956 & 0.940 \\
    conformal coverage    & 0.912 & 0.906 & 0.914 \\
    scale factor $\hat q$ & 1.48  & 1.37  & 1.51  \\
    \hline
  \end{tabular}
\end{table}

\section{Comparison of posterior architectures}\label{app:architectures}
Table~\ref{tab:baselines} compares the posterior architectures on the accuracy and calibration of the posterior mean. Table~\ref{tab:arch} adds the two quantities that Table~\ref {tab:baselines} omits, both under the same per-channel protocol and the same evaluator: the accuracy of the best of 400 posterior draws, and the test negative log-likelihood of the full predictive distribution. The sample oracle differs from the pooled-ensemble parameter oracle of section~\ref{sec:methods}, which selects among the 25 mixture components. Here, each model is scored by drawing 400 samples from its own posterior, so the column is comparable across architectures with no pooling components.

The MDN ensemble, the deep ensemble and the cINN agree to within $0.003$ in mean $\Rsq$, but they differ in what their posteriors contain. The best of 400 draws reaches an oracle $\Rsq$ of $0.999$ for the two multimodal models and $0.996$ for the unimodal deep ensemble. The gap is small because the best of many draws lies close to the truth even for a single Gaussian. The difference between the models is that only the multimodal ones can place mass on the second solution of section~\ref{sec:degeneracy}. The cINN attains the best negative log-likelihood by $0.72$ nats, as a free-form flow is not restricted to five components. The cVAE is the exception in accuracy, losing $0.27$ in $\Rsq$ on $\azero$. It was trained on the same budget and with no architecture search of its own, so we do not claim that the gap is inherent in its model class. Empirical coverage is near-nominal for all five models except MC-dropout, which overcovers markedly. We keep the MDN ensemble because its explicit mixture components are what the mode analysis of sections~\ref{sec:degeneracy} and~\ref{sec:reducing} reads the degeneracy off, and because it trains roughly ten times faster than the flow. 

\begin{figure*}[h]
  \centering
  \includegraphics[width=\textwidth]{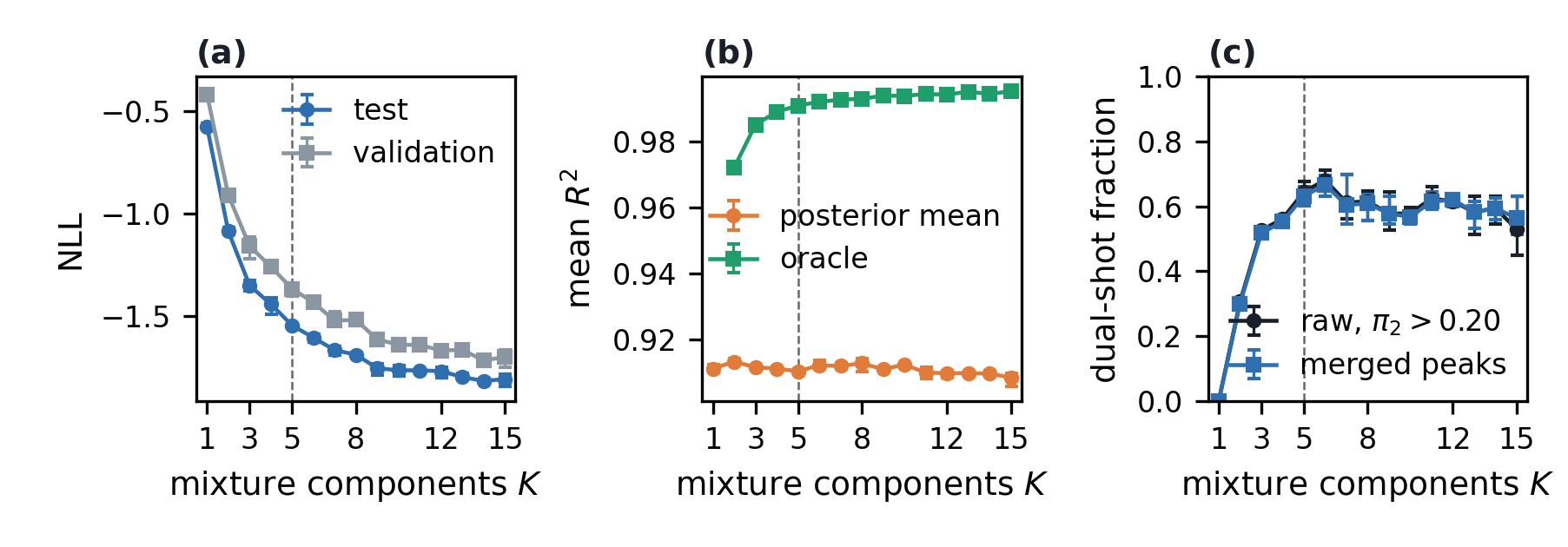}
  \caption{Mixture-component sweep, $K=1$ to $15$, three seed bases each, mean with the min-to-max range (test set, $n=9846$, per-channel noise). (a) Test and validation NLL. (b) Mean $\Rsq$ of the posterior mean and of the oracle. (c) Fraction of dual shots, counted by the second component weight and by merged peaks. The dashed line marks $K=5$.}
  \label{fig:ksweep}
\end{figure*}

We chose $K$ with a sweep from $1$ to $15$, same seeds, same protocol (figure~\ref{fig:ksweep}). The $\Rsq$ of the posterior mean is flat over the whole sweep. Even one Gaussian reaches it. In a dual shot, the posterior mean lies between the two branches for any $K$. So $\Rsq$ cannot tell the mixture from a single Gaussian. The oracle $\Rsq$ and the dual-shot fraction stop changing at $K=5$. The test NLL still improves after that, but slowly. The extra components split existing branches. They do not find new ones. We did not try full covariance matrices. From $K=5$ to $K=15$, the NLL gains less than the distance to the cINN in Table~\ref {tab:arch}. So the component count is not what limits the MDN. The cINN has the better likelihood and reaches the same accuracy and the same branches, so the conclusions of sections~\ref{sec:degeneracy} and~\ref{sec:reducing} do not depend on this choice.

\begin{table}[h]
  \caption{Posterior architectures under the per-channel protocol of table~\ref{tab:noise}, scored with the common sample-based evaluator (400 draws per shot). Mean $\Rsq$ of the posterior mean; oracle $\Rsq$ of the best of the 400 draws; test negative log-likelihood of the full predictive distribution (lower is better); mean empirical coverage of the $68\%$ and $95\%$ intervals. The five rows are the posterior models of Table~\ref {tab:baselines}; the flat-noise comparator, kNN and linear regression are omitted.}
  \label{tab:arch}
  \centering
  \small
  \begin{tabular}{lccccc}
    \hline
    Model & mean $\Rsq$ & oracle $\Rsq$ & NLL & cov68 & cov95 \\
    \hline
    MDN ensemble  & 0.910 & 0.999 & $-1.91$ & 0.70 & 0.95 \\
    deep ensemble & 0.913 & 0.996 & $-0.81$ & 0.67 & 0.95 \\
    cINN          & 0.911 & 0.999 & $-2.63$ & 0.69 & 0.95 \\
    cVAE          & 0.760 & 0.998 & $-1.35^{\dagger}$ & 0.66 & 0.96 \\
    MC-dropout    & 0.902 & 0.996 & $-0.00$ & 0.85 & 0.99 \\
    \hline
  \end{tabular}
  \\[2pt]
  \footnotesize $^{\dagger}$ mean per-member IWAE-64 bound, not an exact likelihood. It is an upper bound on the model's NLL, so it penalises the cVAE.
\end{table}

\section{Noise-sensitivity of the diagnostic channels}\label{app:noise}
We rescale the noise half-width of one diagnostic channel at a time by factors of 0, 0.5, 2, or 4, keep the other channels at their nominal values in Table~\ref{tab:noise}, and re-evaluate the proposed model without retraining. The RMS divergence, the bunch length $\sigma_z$ and the charge $\Q$ dominate: quadrupling the noise of any one of them costs between $0.078$ and $0.107$ in mean $\Rsq$, against $0.010$ to $0.038$ for the remaining six channels. Removing the noise of a single channel entirely never gains more than $0.007$. The model is thus limited by the joint precision of these three channels. Table~\ref{tab:noisesens} gives the expected gain from improving any one of them.

\begin{table}[h]
  \caption{Mean $\Rsq$ of the proposed model when the noise half-width
  of a single channel is scaled by the given factor. The all-nominal
  mean $\Rsq$ is $0.911$.}
  \label{tab:noisesens}
  \centering
  \small
  \begin{tabular}{lcccc}
    \hline
    Channel & $\times 0$ & $\times 0.5$ & $\times 2$ & $\times 4$ \\
    \hline
    $\xof$              & 0.911 & 0.911 & 0.909 & 0.900 \\
    $\Emed$             & 0.915 & 0.912 & 0.906 & 0.887 \\
    $\dEmad$            & 0.914 & 0.912 & 0.907 & 0.892 \\
    $\Q$                & 0.917 & 0.914 & 0.896 & 0.833 \\
    $i_{\mathrm{peak}}$ & 0.914 & 0.912 & 0.906 & 0.887 \\
    $\epsilon_{n,x}$    & 0.914 & 0.912 & 0.905 & 0.882 \\
    $\sigma_y$          & 0.915 & 0.913 & 0.903 & 0.873 \\
    $\sigma_z$          & 0.917 & 0.915 & 0.894 & 0.828 \\
    divergence (RMS)    & 0.918 & 0.916 & 0.890 & 0.804 \\
    \hline
  \end{tabular}
\end{table}

\section{Conditioning on \texorpdfstring{$\pone$}{p1}: a cascade attempt}\label{app:cascade}
The degeneracy couples $\azero$ and $\pone$. We test whether a cascade that first predicts $\pone$ and then conditions the $\azero$ regression on it recovers the lost accuracy. This test predates the per-channel noise model and was run with a flat $20\%$ feature noise and an exact $\xof$, so its levels are below those of the body, and only the comparison in the appendix is meaningful. Conditioning on the true $\pone$ raises the $\azero$ $\Rsq$ from $0.668$ to $0.875$. Conditioning on the predicted $\pone$ reaches $0.659$, no better than the baseline. The gain is therefore bounded by the accuracy of the $\pone$ prediction itself. The oracle value of $0.875$ measures the size of the degeneracy. It does not indicate that a better architecture could reach it.

\section{The paired particle-in-cell campaign}\label{app:pic}
We test the second-working-point argument of section~\ref{sec:twowp} directly on PIC beams, by applying the frozen ensemble of section~\ref{sec:methods} to a dedicated campaign of 100 \textsc{Smilei} simulations at 45 machine settings. At each setting $\azero$, $\xof$ and $\cNtwo$ are held fixed, and $\pone$ is raised by the known factor $r = 2.1$, so the second shot samples the same setting at a higher plasma density; ten of the settings were additionally run at an intermediate factor of $1.55$, which the analysis below does not use. The runs use the same envelope solver as the training library. Both working points are injected in 38 of the 45 pairs.

The campaign uses the same input deck as the training library, so no pair is excluded as out-of-distribution. As a check, the forward surrogate evaluated at the true inputs gives a median $\chi^2$ of $1.04$ for the lower-density shots and $2.73$ for the higher-density ones, within the $0.99$ to $2.89$ range of the training library's test set across the same pressure range.

Each pair is evaluated under 64 independent draws of the per-channel noise of table~\ref{tab:noise}, and four arms are compared. The \emph{single} arm is the posterior $p(\theta \mid y_1)$ of the lower-density shot. The \emph{two-shot} arm is the normalised product $p(\theta \mid y_1)\,p(\theta \mid y_2)$, which is the posterior of the pair when the two measurements are conditionally independent given $\theta$ and the prior is flat on the scanned box. Both factors are Gaussian mixtures, making their product available in closed form. We then transform the posterior of the second shot into the coordinates of the first using the exact $1/r$ rescaling of $\pone$. The \emph{repeat-shot} control replaces $y_2$ with a second noise draw of the same lower-density beam, which supplies the same averaging without a density change. The fourth arm, \emph{high-density single}, is $p(\theta \mid y_2)$ alone, which separates the value of changing the density from the value of operating at the higher one.

Table~\ref{tab:picpairs} bounds what can be read from the campaign. The central $68\%$ intervals are close to calibrated on these beams on both arms, with coverages between $0.62$ and $0.68$. The RMSE ratios separate the two steps of section~\ref{sec:twowp}: on $\azero$ and $\cNtwo$ most of the gain of the pair comes from the pairing itself, which lies above the factor of $1.5$ that the campaign resolves, while a single shot moved to the higher density reaches only $1.17$ and $1.12$. Only on $\pone$, where the density change buys the most, do the two steps contribute about equally. The single-shot error is only partly a bias, and the repeat-shot control gains a factor of $1.09$ to $1.12$; the gain from the density change is measured against this baseline of about $1.1$. A systematic error common to the two shots, such as a drift in the laser energy between them, would violate the conditional independence assumed by the product. The two simulated shots share no such term. The simulations establish the direction and approximate magnitude of the gain. An experiment would still need to verify that the two measurements are sufficiently independent.

\begin{table}[h]
  \caption{Diagnostics of the 38 pairs, over 64 noise draws each. Empirical coverage of the central $68\%$ intervals on the single and two-shot arms; magnitude of the bias as a fraction of the single-arm RMSE; RMSE of each remaining arm relative to the single arm, where a larger number is a smaller error; and the pair relative to the high-density single shot, which isolates the pairing itself.}
  \label{tab:picpairs}
  \centering
  \small
  \begin{tabular}{lccc}
    \hline
    & $\pone$ & $\azero$ & $\cNtwo$ \\
    \hline
    cov68, single arm                       & 0.66 & 0.62 & 0.63 \\
    cov68, two-shot arm                     & 0.67 & 0.62 & 0.68 \\
    $|\mathrm{bias}|/\mathrm{RMSE}$, single & 0.38 & 0.34 & 0.15 \\
    RMSE gain, repeat shot                  & 1.12 & 1.09 & 1.12 \\
    RMSE gain, high-density single          & 2.13 & 1.17 & 1.12 \\
    RMSE gain, pair                         & 4.05 & 2.59 & 2.95 \\
    pair over high-density single           & 1.90 & 2.21 & 2.64 \\
    \hline
  \end{tabular}
\end{table}

\section{Virtual-diagnostic drift run}\label{app:drift}
\begin{figure*}[h]
  \centering
  \includegraphics[width=\textwidth]{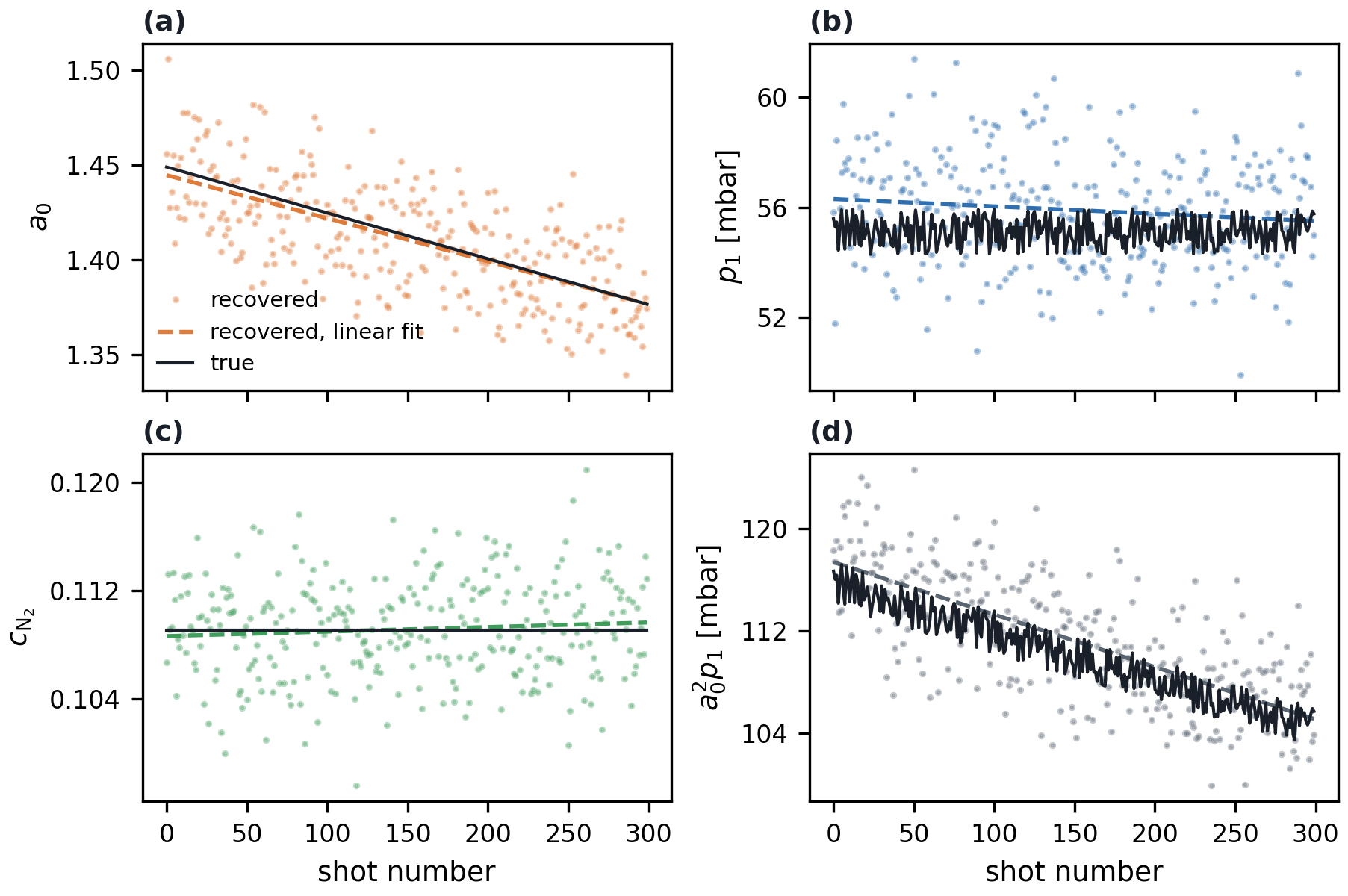}
  \caption{Simulated drift run of 300 shots at a fixed working point. (a) $\azero$ drifting down by $5\%$, (b) $\pone$ with shot-to-shot jitter of $\pm 1.5\%$, (c) constant $\cNtwo$ and (d) the combination $\azero^2\pone$. Solid lines show the true values, points the posterior means of the proposed model under per-channel noise, and dashed lines linear fits to the points.}
  \label{fig:drift}
\end{figure*}

We simulate a run of 300 shots at one working point of the test set, with $\pone = 55.1$~mbar, $\azero = 1.45$ and $\cNtwo = 0.109$. Over the run, $\azero$ drifts down by $5\%$. The pressure $\pone$ jitters from shot to shot within $\pm 1.5\%$. The nitrogen concentration $\cNtwo$ and the focal position $\xof$ stay fixed. The clean-trained forward surrogate gives the beam of each shot. The per-channel noise of Table~\ref{tab:noise} is applied, and the proposed model recovers the inputs (Figure~\ref{fig:drift}).

We fit a linear trend to each of the 300 shots for the true and recovered quantities. The recovered $\azero$ follows the drift with $94\%$ of the true slope. The trend in the recovered $\cNtwo$ stays below $1\%$. The recovered $\pone$ drifts by $-1.4\%$, although the true $\pone$ has no trend. For $\azero^2\pone$, the recovered change is $-11.1\%$ and the true change is $-10.2\%$.

\bibliographystyle{iopart-num}
\small
\bibliography{references}

\end{document}